\documentclass[a4,50pt]{amsart}%
\usepackage{amssymb}
\usepackage{pdflscape}
\usepackage{amsmath}
\usepackage{mathrsfs}
\usepackage{amsfonts}
\usepackage{graphicx}%
\usepackage[all]{xy}
\usepackage[colorinlistoftodos]{todonotes}
\usepackage[colorlinks=true, allcolors=purple]{hyperref}
\theoremstyle{plain}

\numberwithin{equation}{section}
\begin{document}

\title[Schr\"odinger's Cat]{Schr\"odinger's Cat}
\author{M.~F.~Brown}
%\institute{M. F. Brown  \\ \email{m135brown@gmail.com}          }
\date{Began circulation August 2019. This version: September 2022 corrects eq.~(7.8) in January 2021 version}
\begin{abstract}
The following
article began as a contribution to an unfinished paper \cite{Gil09}, specifically with regards to constructing a model of the Geiger Counter. What is presented here is essentially an example of Belavkin's quantum filtering theory \cite{Be92} applied to the infamous Schr\"odinger's Cat problem. Although the basic mathematical framework is the same as that presented in the exposition \cite{Bel02} the explicit details of the construction are different and more natural from the point of view of the underlying physics.

The basic idea here is that observation (or one's experience) is fundamental and the `atomic world' is postulated as the source of such observation. Once this source has been inferred to exist one may attempt to explicitly derive its structure in such a way that the observation itself can be reproduced. And so here is a purely quantum mechanical model of observation coupled to its supposed source, and the observation itself is realised as a projection of this quantum system.

%Alternatively, one may view this model of the Schr\"odinger's Cat problem as one that gives rise to the exponential decay of an excited state of a quantum system, `atom', into its ground state. In particular, the atom does not decay in the absence of observation. %Here it is supposed that Schr\"odinger's cat is an apparatus that is coupled to the atom in such a way that if the cat is observed to be dead (assuming that it was alive at the beginning of the experiment) then the atom is inferred to have certainly decayed. However, if the cat is observed to be alive then one cannot extract any useful information about the system if the initial atomic state is unknown.

%Conventionally, observations of the cat being either dead or alive are the only possible ones. But that will not give rise to a marginal atomic state that is continually decaying. Here the observation of a \emph{dying} cat is introduced so that the apparatus has 3 degrees of freedom even though the atom has only 2. The \emph{dying-cat} state is essentially the occurrence of a `click' if the cat were replaced with a Geiger counter.
%In fact one may simply use 2 apparatus degrees of freedom: \emph{dying} and {not dying}, but this does not give such sensible physics.
\end{abstract}
\maketitle
\tableofcontents
\section{Introduction}

We observe, feel, whatever... but where does this sensation come from? The logical way to tackle this problem is to postulate the existence of a `source' that gives rise to our experience. So far, physics has shown us that such a source appears to be `quantum', and perhaps our experience too. But there is a little more to it than this, as we shall soon see. In what follows we shall simply refer to such a source as a `physical system' or `quantum object', eventually settling on the word `atom'. %\footnote{One might ask what is an atom? and there are two things to distinguish: One is that an atom is a source of observation from which we infer a structure of atomic transitions in order to explain our experience of colour, say. The other is that the atom is the observation itself, and the underlying source is something else...}.

Alternatively, if a physical system is believed to have objective structure then it cannot stand alone outside of the context of being `under observation'. %The idea of a system existing in the absence of whosoever is perceiving such a thing is speculation.
If it is only of sense to postulate a physical system that is under observation, it follows that any such physical system is open since it must be communicating, or sharing information, with the observer. %The Universe is the whole, and any system may be regarded as a sub-system of the Universe. By definition, the Universe cannot be opened - if it could then what one previously thought to be the Universe was in fact only a sub-system of it.
%The idea here is that observation, in its most general form, is asking a system questions  and receiving answers back from it. %It is plausible that this interaction has a symmetry?
Thus the minimal structure of a closed system should  be both a quantum object and an observer, or observation. A general case of two mutually observing quantum systems has also been developed in  \cite{Bro13} following the work on `double products' pioneered by Hudson, in, for example \cite{Hud09}.

The next important point is that sensible observation requires a causal structure: a means by which the events forming an observation can be ordered. Any sense of `what will happen next?' requires a statistical predictability, of future events given prior events, which will ultimately require that an observed time-like trajectory  $\{N_s:s\leq t\}$ must commute with itself at different times, as well as with any quantum property $X$ of the system being observed \cite{Bel94}.
%This process is obviously causal, otherwise it is nonsense. %Further, this object-observer interaction is assumed to asymmetric in so far as  the dual system*=observer is not considered here - see double products.
%Further, this system-observer interaction is regarded as a universal mechanism applicable to any pairs of subsystems of the Universe, where either component of the pair may be regarded as the system and the other the observer.

One may think of the object-observer interaction as a sort-of questioning and answering type of interplay between these two systems,
%The phrase `asking a system questions' is metaphorical and meant in the most general way.
and in the spirit of physics one may regard such `questioning' as the application of force by the observer upon the system, and the `answering' as the reaction of the system upon the observer.% Already we have diverged from Newton's thinking as an equal and opposite reaction would not suffice as a means to extract non-trivial information from a system.
%At present the most satisfactory mathematical framework for extracting information about subsystems of the Universe is that of Quantum Filtering as it generalizes the mathematics of observation of general random processes to quantum systems.

\section{Open quantum systems}
In the \emph{interaction picture} the state of an open quantum system is understood to evolve according to the Lindblad equation
\begin{equation}
\frac{\partial}{\partial t}\rho(t)=-\frac{1}{2}\big\{L^\ast L,\rho(t)\big\}+L^\intercal\rho(t)L^\dag\label{2}
\end{equation}
where $L^\intercal=(L_1,L_2,\ldots)$ is a row of bounded operators $L_k\in\mathcal{B}(\mathfrak{h})$ acting on a Hilbert space $\mathfrak{h}$. The algebra $\mathcal{B}(\mathfrak{h})$ is equipped with the usual Hermitian involution denoted by $\dag$, and the adjoint of $L^\intercal$ is denoted by $L^{\intercal\ast}=L^\dag$ which is a column of operators $L^\dag_k$. The transposition given by $\intercal$ shall be referred to as `outer transposition'. It is understood within the framework of QSC (\emph{Quantum Stochastic Calculus}) \cite{Bel92,HuPa84} that (\ref{2}) is in fact the marginal dynamics of a unitary quantum stochastic evolution where the aforementioned quantum system is composed with a field of stochastic noise which drives the dynamics of the quantum system. This was of course the idea behind Lindblad's derivation.

The index $k$ on the operators $L_k$ labels the degrees of freedom of the noise and these degrees of freedom are often interpreted, equivalently, as the degrees of freedom of an environment within which the quantum system sits. In fact one can  understand this environment as an observer % (this naturally arises from QSC) % and the resulting dynamical interpretation of the quantum measurement projection postulate, where the time-continuous dynamics of a quantum system under observation gives rise to the same marginal dynamics (\ref{lin}).
and then the index $k$ would label the degrees of freedom of the observations being made. This would mean that the observations that one makes would drive the dynamics of the source responsible for them, ultimately affecting future observations. Or, put another way, whatever abstract question the observer asks of the quantum system\footnote{One could phrase this more generally as: whatever concepts the observer forms about the universe.} will result in a set possible answers indexed by $k$. And so it comes to pass that the marginal dynamics of the quantum system is governed by the possible answers to the question you asked.

%%%%%%%%%%%%%%%%%%%%%%%%%%%%%%%%%%%%%%%%%%%%%%%%%
%%%%%%%%%%%%%%%%%%%%%%%%%%%%%%%%%%%%%%%%%%%%%%%%%
%However, such marginal dynamics is not a realistic evolution of the quantum system because it is purely speculative in the absence of actual observations. One could argue that it could be like this if the system was unsure what answer to give, but `no answer' could also be encoded into the observing `apparatus' as an extra degree of freedom. One of the difficulties here is determining %in a realistic way what the operators $L_k$ should be. One's best attempt might be: \emph{``if the answer $k=1$ means that the system is in a state $\psi_1$ then $L_1:\psi\mapsto\psi_1$ up to some scalar factor''}. Following this, one could then attempt to perform a different kind of measurement/observation of the system to try and conclusively determine that \emph{``the state certainly was $\psi_1$''}. And such is the nature of precision quantum experiments.
%%%%%%%%%%%%%%%%%%%%%%%%%%%%%%%%%%%%%%%%%%%%%%%%%%
%%%%%%%%%%%%%%%%%%%%%%%%%%%%%%%%%%%%%%%%%%%%%%%%%

In what follows, a quantum system consisting of a single quantum bit shall be considered as the source. It shall be thought to be in an arbitrary state $|\psi\rangle$ and believed to decay permanently into a ground state $|g\rangle$ of its Hamiltonian. In an attempt to infer this decay the `atom' is supposed to be prepared in such a state and then coupled to Schr\"odinger's cat\footnote{This is a rather synthetic quantum experiment type of setup and one might like to have in mind something more like: the life of the cat is determined by the existence of a quantum bit with which the cat is naturally coupled to. This quantum bit was not `prepared' in a laboratory, it is simply an intrinsic part of the universe responsible for the fate of the cat. Its existence is postulated to explain the fate of the cat.}. At this stage both the atom and the cat are regarded as `quantum' in so far as they shall be modelled using the quantum formalism (wave-functions and observables) and the two shall be coupled throughout an interval of time by a unitary operator $U_t$, forming an entangled quantum superposition.

\section{A continual decay}\label{sec3}
So we are modelling an atom as a 2-level system in an arbitrary initial state $|\psi\rangle=\alpha|g\rangle+\beta|e\rangle$
where $|g\rangle$ is the ground state and $|e\rangle$ is the excited state forming an orthonormal basis for a Hilbert space $\mathfrak{h}$
containing such state-vectors $|\psi\rangle$. If the atom is considered to decay from its initial state into the ground state then the expected (averaged) behaviour of this decay process may be regarded to have the form
 \begin{equation}
 \rho(t)=|\psi\rangle\langle\psi| e^{-\nu t} +|g\rangle\langle g|(1-e^{-\nu t})\label{1}
 \end{equation}
 in the interaction picture,
 and if one feels that this is not general enough then the decay rate $\nu$ may be replaced with a time dependant decay rate\footnote{In what follows generalising
  to a time dependant decay rate is not a problem.}.

  The next thing to do is to consider a differential equation that corresponds to such a decay, and one may easily check that (\ref{1})
   solves the Lindblad equation (\ref{2})
with $L^\intercal=\sqrt{\nu}(-J^\dag J,J)$ %recall that $L^\ast=L^{\intercal\dag}$ with $\dag$ denoting the usual Hermitian Involution in $\mathcal{B}(\mathfrak{h})$,
and $J:=u|g\rangle \langle e|$, where $u=e^{-\mathrm{i}\varepsilon t}$ is just a scalar unitary arising from the interaction picture. %in fact $u$ may simply be ignored as it has no important role in what follows.
Note that we could have simply differentiated (\ref{1}) to
  obtain a differential equation, but the benefit of (\ref{2}) is that it is given entirely in terms of operations on $\rho(t)$.
   This is good because it sheds light on the existence of the fundamental operation, $J$, corresponding to the atomic decay\footnote{It would be nice to have had an even simpler Lindblad equation with $L$ having only one degree of freedom instead of two so that $L=\sqrt{\nu}J$. This would
be ok, as we would have something similar to (\ref{1}) except that the off-diagonal terms of $\rho$ would decay more slowly
so that there is no longer a uniform decay of $|\psi\rangle\langle\psi|$. Further, to obtain a unitary dilation in this case would require a more complicated procedure than that resulting in the form (\ref{s}).}.

The Lindblad equation is often referred to as the `master equation' in quantum mechanics, describing the Markovian evolution of an open quantum system; the `atom' here is an open system so everything appears to be in order. If a system is open one may realise a bigger closed system by introducing an \emph{environment}. This is basically Stinespring dilation, and the unitary dilation of Lindblad dynamics is in fact a special case of the more general QSC. %Sadly, this is unknown in most of the world of quantum physics.
In the case of this qubit atom equation (\ref{2}) is very basic and thus serves as an excellent introduction to the subject; it is also a very important example.

This basically means that the Lindblad equation can be realised as an equation describing the \emph{marginal} deterministic dynamics of a more general quantum stochastic process,
and such a process is just the joint (entangled) evolution of our atom with an environment. What this environment {is} remains to be discussed.
Now, the dilated dynamics incorporates, precisely, the idea that the atom's decay is in fact spontaneous, and the marginal state (\ref{1})
 is simply an averaged-out form of the evolution conceding to a lack of knowledge of when the spontaneous decay actually occurs.
 On the one hand, the supposed underlying stochastic evolution of the atom is no more than a Stinespring-type dilation, with the marginal state
  $\rho(t)$ corresponding to a continual application of a quantum channel. But on the other hand, it makes way for a philosophically sound dynamical
  description of quantum evolution incorporating observation.

\section{A spontaneous decay}
Notice that (\ref{1}) may be written more generally as
 \begin{equation}
 \rho(t)=|\psi\rangle\langle\psi|\; \mathbb{P}[Z> t] +|g\rangle\langle g|\;\mathbb{P}[Z\leq t]\label{3}
 \end{equation}
 where $Z$ is a random variable representing the time at which the atom spontaneously decays.
 Here we consider $Z$ to be distributed exponentially with parameter $\nu$. All (\ref{3}) has achieved is to recognize that the decay is spontaneous
 in some kind of bigger picture (it is a `bigger' picture because we have introduced an additional random variable corresponding to a \emph{time coordinate}
 variable). So let's make this more rigorous. To do this let's recall the Lindblad equation (\ref{2}) and its main ingredient $J$.

 The operator $J$ induces the decay of $|\psi\rangle$ into the ground state, so its action is synonymous with the spontaneous decay of the atom. Of course $J$ is not unitary, but the simplest unitary dilation of it is
\begin{equation}
S=\left(
    \begin{array}{cc}
      JJ^{\dag}  & J^{\dag} \\
      J & J^{\dag}J \\
    \end{array}
  \right).\label{s}
\end{equation}
Such a unitary dilation is necessary otherwise our system will not be closed.
What we have essentially done here is introduce a second system with two degrees of freedom (note that our Lindblad operator $L$ had two degrees of freedom, and is indeed related to $S$, see appendix \ref{ap2}).
Further, we have introduced a product-type composition of the atom with this auxiliary system (this auxiliary system is a fundamental building block for the environment). %In the spirit of the mathematics of algebra we are also required to
We shall represent this auxiliary system by a Hilbert space $\mathfrak{k}$ and we shall work in the standard basis $|0\rangle=(1,0)^\ast$ and $|1\rangle=(0,1)^\ast$,
%\begin{xrem}
%Notice that the usual Hermitian conjugation is given by $\dag$ on $\mathfrak{h}$ and $\ast$ on $\mathfrak{k}$. This is indeed intentional because in the
%general theory there is a requirement for more intricate involutions on $\mathfrak{h}\otimes\mathfrak{k}$.
%\end{xrem}
which is determined by the chosen form of $S$. In this case the `initial' state of the auxiliary system is $|0\rangle$ as it is actually the isometry $F:=S|0\rangle$ that is the key feature here for driving the dynamics (\ref{2}). With respect to this isometry the \emph{spontaneous} transformation of the atomic state at any time $t$ is
\begin{equation}
\rho(t)\mapsto \texttt{Tr}_{\mathfrak{k}}\big[F\rho(t)F^\ast\big]\equiv F^{\dag\ast}\rho(t) F^\dag=J J^{\dag}\rho(t) J J^{\dag}+ J \rho(t) J^{\dag} ,
\end{equation}
and $F^{\dag\ast} \rho(t) F^\dag=|g\rangle \langle g|$ no matter what $\rho(t)$ is. This is of course a quantum channel, and notice that $|g\rangle \langle g|$ is invariant under this transformation. Roughly speaking, this is our continuously applied quantum channel\footnote{This is not quite true. Note that a quantum channel is a quadratic form, but the actual quantum channel that describes the Lindblad dynamics is
 a \emph{pseudo} quadratic form. This basically means that the Stinespring dilation of an open system's stochastic dynamics is represented on a
  pseudo Hilbert space \cite{Bel92}; that contains non-zero vectors with a zero-valued inner product!
  That is the Belavkin Formalism of quantum stochastic calculus, about which some basic details have been included in the appendices \ref{ap2}.} mentioned at the end of Section \ref{sec3}, as we shall see.

The action of $S$ on the dilated system $\mathfrak{h}\otimes\mathfrak{k}$ will of course occur at a random time $Z=z$ in accordance with (\ref{3}),
 and such exponentially distributed $Z$ will in fact emerge from a consideration of repeated random actions of $S$ in the form of a quantum Poisson process\footnote{Perhaps this does not seem necessary  as a single application of $S$ gets the job done (i.e. decays the atom).
But we are really trying to illustrate a general formalism here.}. This means that we will be considering chains,
  $\vartheta$, of random times when $S$ acts, and the only way that we can define a repeated action of $S$ is to have a copy of $\mathfrak{k}$ at each such random time $z\in\vartheta$.
Moreover, we shall see below that a repeated action of $S$ not only gives us all the structure that we are looking for but also gives us no unwanted structure.
%However, if one raises the question of `why did the atom spontaneously decay?' then the only sensible answer resulting in progress is that `something triggered the decay.' Thus it is natural to consider an incoming flux of \emph{event triggers}, rather than a single isolated trigger for the decay. That is simply
Thus we have a composition, or string, of copies of the auxiliary system $\mathfrak{k}$ all in the state $|0\rangle$ and each copy of $\mathfrak{k}$ is ultimately there to carry a copy of $F=S|0\rangle$, the operator that generates the dynamics (\ref{2}). %What we are actually talking about is a repeated action of $F$ and this is really the only way that the repeated action can be conceived. %The first is $S^n$ if we consider the action $n$ times. When considering all possible values of $n$ we would end up with a Fock space of scalar `particles' composed with $\mathfrak{h}\otimes\mathfrak{k}$.

\section{Preparing for measurement}

%We are now in a position to claim that there is some kind of `hidden' system (microscopic or hidden from our direct observation)  about which we can ask
%questions - with the intention of explaining our observations. Something is there causing us to have non-trivial observations. % and our observations was cause it to evolve
Experiments in physics are trying to set up very controlled environments so that very precise statements can be made about the origins
 of our observations (LHC et c.). But the current mathematical formalism (QFT) does not properly account for and distinguish between the origins of observation and the observation itself.
 In this article our `decaying atom' is purely hypothetical, but its purpose is to represent a basic source of observation. One may regard oneself as the observer here and then any such thing as this `atom' should be arrived at as a concept through which we attempt to explain and understand our observations. Further, through inferring this source of observation we would then hope to develop knowledge of it, through which we might hope to anticipate and improve our future observations; improve our experience in general.

So we shall accept that there is a hidden atomic world from which our observations arise and we are probing it with  questions
corresponding to the observations we have. We might ask \emph{``has the atom decayed?''} and then we  observe, by virtue of a device such as a
 Geiger counter or Schr\"odinger's cat, an answer. %It should now be more obvious that the auxiliary system $\mathfrak{k}$ and its second quantization have something very fundamental to do with our observation; our observation ultimately results in knowledge about the atom. Further,
Whatever apparatus we are using to make our observations (one's self or an instrument) must couple to the system that we are questioning in the manner of our
 previously discussed `auxiliary' system. Further, the results produced by the apparatus (the observations that we can make) must correspond to its
  degrees of freedom\footnote{Indeed one might expect to have at least two degrees of freedom corresponding to some kind of `yes' or `no' outcomes.}.
   %\footnote{  One must take care in interpreting the meaning of the mutually exclusive apparatus states $|0\rangle$ and $|1\rangle$ corresponding to the basic observations that may be made. In terms of Schr\"odinger's cat the first corresponds to a `steady state' of the cat, which may either be alive or dead, it doesn't matter which, whilst the second corresponds to a `transition' from alive to dead. %even if death not observed but instead one instant alive and the next dead, 1 encodes this. This basically corresponds to encoding an aspect of physical reality into the mathematics.}.
  So each copy of $\mathfrak{k}$ should be interpreted as the measurement apparatus at an instant of time. This basically means that the
   realisation of us probing (questioning) the atom is the action of the operator $F$, and the environment \emph{is} the observer\footnote{To be clear: we first considered an open atomic system, then dilated it with an auxiliary system, then understood ourselves (or some instrument) as  that auxiliary system. But the real world is the reverse of this: there is us (or an instrument) that is observing, and an atomic system is inferred as the source of that observation.}.

Now we have all the criteria to consider our observation of Schr\"odinger's cat.  The idea is this: the cat is serving as our observation and the observation is only regarding whether the cat is alive or dead. We assume the cat is alive at time $0$ and if the cat is still alive at time $t$ then we can't infer anything about the source of observation because nothing has happened. But if the cat is observed to be dead at time $t$ then we can infer that there is some origin of that observation. In particular, it is the actual death of the cat that leads us to postulate some kind of atomic event: the atomic decay. There is a subtle  point here. Usually, people work with the cat states `alive' and `dead' but this will not give rise to the atomic decay, nor will it allow a transition from `alive' to `dead' leaving us stuck with some kind of Zeno's paradox type scenario.
%\cite{MisS77}.
Here these issues are easily overcome by working with the cat states `dying' and `not dying', respectively $|1\rangle$ and $|0\rangle$ instead\footnote{This all comes down to the question of how one can encode aspects of the real world into mathematical frameworks.}.

Let's suppose that the experiment occurs on an interval of time $[0,r)$, then we shall parameterise evolution on this interval with $t$. The auxiliary system $\mathfrak{k}$ now represents the cat states described above, and the operations $S$ represent spontaneous interactions between the atom and the cat. These interactions are then postulated to be the origin of both the atomic decay and the cat's death, and our observation of the cat's death infers the atomic decay as the cause\footnote{In an equal setting  the cat's death can of course also be thought of as the source of the atomic decay \cite{Bro13}.}.

The operator $F$ has components $F_0=|g\rangle\langle g|$ and $F_1=J$ and during the experiment
there could be any number of actions of $F$, so we must represent this dynamics on the Hilbert space $\mathscr{H}=\mathfrak{h}\otimes\mathscr{F}$
where $\mathscr{F}$ is a Guichardet-Fock space (see the Appendices \ref{ap1}) over $[0,r)$ generated by $\mathfrak{k}$. The corresponding state in this Fock space would be a coherent state
$\Phi=|0\rangle^\otimes \sqrt{\nu}^\otimes e^{-\frac{1}{2}\nu r}$,
where the constant function $\nu$ is the rate of spontaneous interactions between the atom and the cat.
% and one may regard $\sqrt{\nu}$ as a coupling strength (of the atom to the apparatus).
A unitary operator $U=S^{\odot}$ (given by a semi-tensor product $\odot$ (\ref{odot}) of the $S$ operators  so that one at a time they act upon $\mathfrak{h}$ and the next copy of $\mathfrak{k}$) describes
the full interaction between the atom and the cat on $[0,r)$, coupling the two together and  resulting in the \emph{prepared state} $\Psi=
U(|\psi\rangle\otimes\Phi)$ which is of course an entangled quantum state holding all the information about any future observations that may be made.

The Hilbert space $\mathscr{H}$ admits the \emph{future-past} decomposition $\mathscr{H}_{[t}\otimes\mathscr{F}^t$ with respect to a present time $t$ where
 $\mathscr{H}_{[t}=\mathfrak{h}\otimes\mathscr{F}_{[t}$, and
$\mathscr{F}^t$ and $\mathscr{F}_{[t}$ are respectively Fock spaces over $[0,t)$ and $[t,r)$. With this in mind we will actually be considering entanglement up to time $t$:
\begin{equation}
\Psi_t=U_t\big(|\psi\rangle \otimes \Phi\big),
\end{equation}
where $U_t=I_{[t}\otimes U^t$ is a quantum stochastic Poisson process in our example here. %, but don't worry about that.
This state may be evaluated on any random chain of coordinate-times, defined as $\vartheta=\{t_1<\cdots<t_n\}\subset[0,r)$, corresponding to possible instants when the atom interacts
with the cat; which is non-trivial for the part $\vartheta^t:=\vartheta\cap[0,t)$. For any such evaluation $\Psi_t$ simplifies, by virtue of the particular form (\ref{s}) of $S$ for the decaying atom, to
\begin{equation}
\Psi_t(\vartheta) = %\big(\Phi(\vartheta_{[t})\otimes F^\odot(\vartheta^t)\big)|\psi\rangle
%=\big(\Phi(\vartheta\setminus t_1)\otimes F(t_1)\big)|\psi\rangle,
\sqrt{\nu}F|\psi\rangle\otimes \Phi(\vartheta\setminus t_1),\label{Psi}
\end{equation}
so only the first action of $S$ has any effect\footnote{A more realistic observation using the cat is constructed in the sequel. But the construction here is sufficient for illustrating the most important ideas.} (and $t_1<t$ has been assumed so that $t_1\in\vartheta^t$ i.e. so that $\vartheta^t$ is not empty).

\section{Reduction, Expectation and Observables}

Now we shall put all the pieces together and trivially recover the dynamics (\ref{1}) and we shall also look at the observables relevant to this model.
%experiment (the experiment is of course the deduction of atomic decay based on observing  Schr\"odinger's cat (or a Geiger counter)).
First of all note that the marginal density matrix $\rho(t)$ obtained by tracing over the whole Fock space is
\begin{equation}
\rho(t)=\texttt{Tr}_{\mathscr{F}}\big[ \Psi^{}_t  \Psi_t^\ast\big]=%{F_t^{\odot\dag}}^\ast|\psi\rangle\langle\psi|F_t^{\odot\dag}=
 |\psi\rangle\langle\psi| e^{-\nu t} +|g\rangle\langle g|(1-e^{-\nu t})
\end{equation}
which is a good start as it is of course the same as (\ref{1}) (though not surprising); and that is the dynamics recovered.
Note in particular that the exponential decay is a direct consequence of having the Fock space which supports the quantum Poisson process.
Secondly, we must now consider the \emph{true} observables for this experiment.
%It should be quite obvious that
Such observables must simply be operators on the Fock space as that is the space representing the observations. Further, since the observations are being represented in the chosen basis they must be diagonal in that basis. Here the cat (the apparatus) has two degrees of freedom indexed by 0 and 1 so that all {observables} are ultimately made from (tensor) products and sums of the projectors
\begin{equation}
P_0=|0\rangle \langle0|,\quad\textrm{and}\quad P_1=|1\rangle\langle1|.
\end{equation}
That said, one may still consider quantum properties $X=X^\dag$ on $\mathfrak{h}$ and obtain  expectations of $X$ (that are possible answers to the question `$X$'),
 \emph{but to do this} $X$ \emph{must be projected onto the algebra generated by the observables}\footnote{Because that, and only that, is where the observer lives.}.
 This is achieved by a conditional expectation\footnote{Which is the subject of Belavkin's non-linear filtering theory.} and we shall return to this in  Section \ref{fil}.
 Notice that $X$ is referred to as a `quantum property' not an observable.

 The observables that we are looking for should be directly related to what we are observing.
 So one of the most important ones in this case is an operator representing the actual act of counting to one, and only one, since Schr\"odinger's cat can only die once\footnote{This might also be the single click of a Geiger counter.}.
 To this end we shall consider the number operator
\begin{equation}
N_t^1(\vartheta)=\sum_{z\in\vartheta^t}I(\vartheta_z)\otimes P_1(z)\otimes I(\vartheta^z)\label{num}
\end{equation}
where $P_1(z)$ is just $P_1$ identified at the time $z$ and $I(\vartheta_z)\otimes I(\vartheta^z)=I(\vartheta\setminus z)$
is the identity operator on the remaining times in $\vartheta$. Here we have set $\vartheta_z=\vartheta\cap(z,r)$ as distinct from
 $\vartheta_{[z}=\vartheta\cap[z,r)$.\footnote{Remember we are working in the interaction picture, but the free-shift on the domain of $\vartheta$ leaves the interaction operator $S$ invariant here.}  %In the Schr\"odinger picture the present is fixed at the origin  of the coordinate-time so that all atom-apparatus interactions, given by $S$, happen at  $z=0$. This means that in the Schr\"odinger picture $\Psi$ propagates towards the present from the future. This backwards shift of $\Psi$ is generated by the conjugate `clock momentum', $\mu\leftrightarrow\mathrm{i}\frac{\partial}{\partial z}$,  of the coordinate-time  in $\mathscr{F}$, and in the interaction picture this free-shift dynamics is attributed to a forwards shift of the observables and system properties. Thus, the Schr\"odinger operator $S$ takes the form $e^{\mathrm{i}\mu t} S e^{-\mathrm{i}\mu t}=S(t)$ in the interaction picture, which here simply coincides with $S$, and the same goes for $P_1$. All observables and system properties in the interaction picture have the form $T_t^\ast X T_t$ where $T_t=e^{-\mathrm{i}M t}$ is the second quantization of $e^{-\mathrm{i}\mu t}$, and notice that  any quantum property in $\mathcal{B}(\mathfrak{h})$ is invariant under the shift due to its commutativity with $M$.}

%Remember, $\vartheta$ indicates a random chain of instants at which events are considered to take place. These are all just `possibilities' and as such they live
%in the future.

Expectations of general operators $B\in\mathcal{B}(\mathscr{H})$ are given by
\begin{equation}
\mathbb{E}_t\big[B\big]:=\Psi_t^\ast B \Psi_t^{},
\end{equation}
but notice that if $B$ is to be a \emph{predictable property} then it must be compatible with the %experiment's
observables. This requirement is essentially Belavkin's \emph{non-demolition
principle}. Indeed any property $X$ of the atom is predictable
because it commutes with all Fock-space (apparatus) operators. So here is a natural place to introduce the  algebra of  observations $\mathcal{C}_t\subset\mathcal{B}(\mathscr{H}) $
and its commutant $\mathcal{A}_t\subset\mathcal{B}(\mathscr{H})$. The algabra $\mathcal{C}_t$ is the \emph{observer algebra} and it is the %biggest
algebra generated by all relevant compatible observable
processes. %relevant to the experiment,
Whilst the algebra $\mathcal{A}_t$, which contains $\mathcal{C}_t$,  is where all \emph{predictable properties} of the total system live. In
particular it contains all quantum properties of the atom\footnote{To try and make Belavkin's quantum causality clear: suppose we have a free quantum particle with position operator $X(t)$. For $t>s$ can we ask \emph{``what is  $\mathbb{P}[X(t)\in\Delta_t|X(s)\in\Delta_s]$?''} No we cannot. Why? Because $[X(t),X(s)]\neq 0$. is that true? Yes, and to see it note that $\mathbb{P}[X(t)\in\Delta_t]\neq \mathbb{P}[X(t)\in\Delta_t,X(s)\in\Delta_s]+\mathbb{P}[X(t)\in\Delta_t,X(s)\notin\Delta_s]$.}. %\footnote{One might ask `why bother with all this!?' but one of the problems with quantum mechanics is that the so called `observables' of quantum systems don't always commute with themselves at different times. This means that trajectories in quantum mechanics should not be observable! So what do we actually observe when we are doing experiments involving quantum systems?}.

In the case of  the number operator $N^1_t$, counting the number  of deaths of the cat, the expectation at time $t$ is
\begin{equation}
\mathbb{E}_t\big[N^1_t\big]=(1-e^{-\nu t})|\beta|^2
\end{equation}
which begins as zero and increases up to $|\beta|^2\leq1$, which was the initial probability of the atom being in its excited state. So our counting is not expected to
exceed 1, which is of course what we want.

We can also consider another number operator  $N_t^0$.
%counting the number of times that the observer reads the Geiger counter with no click, or peers in the box to see a dead cat.
It is defined as above (\ref{num}) but with $P_0$ instead of $P_1$ and corresponds to redundant observations that preserve the state of the apparatus, i.e. the number of
times that Schr\"odinger's cat is observed to be `not dying' which could mean that it is either dead or alive, but it is not the observation of its death.
This time we of course find that
\begin{equation}
\mathbb{E}_t\big[N_t^0\big]=\nu t -(1-e^{-\nu t})|\beta|^2
\end{equation}
which is just $\nu t$ if the atom is initially in its ground state ($\beta=0$), otherwise this expectation just drops a little below $\nu t$ and then converges
 to it from below. Which makes sense, because $\nu t$ is the expected number of observations of any kind\footnote{Corresponding to the counting $N_t=N_t^0+N_t^1$ of arbitrary events.} and if the event of type 1 is observed then the number
 of events of type 0 will be less than $\nu t$.
%So what is this second system. We have introduced the term `particles' in a rather crude way, so let's think of this as nothing more than terminology. These particles essentially indicate, randomly, the events when the atom experiences a spontaneous transformation.

Note that a random event corresponds to the action of the operator $F=(F_0,F_1)^\intercal$ which does one of two things:
it either transforms the atom into its ground state if it is in any other state, $F_1=J$,   or it preserves the ground state of the atom, $F_0=JJ^\dag=|g\rangle\langle g|$.
 %All of these particles are `prepared' in the state $|\uparrow\rangle$ which has no particular meaning at first. However, under the action of $S$ this particle is transformed into a superposition of $|\uparrow\rangle$ and $|\downarrow \rangle$ states, although all subsequent particles remain in the $|\uparrow\rangle$ state after subsequent actions of $S$.
%events. switch - not activated once atom has decayed.
The former is counted by $N_t^1$  and the latter is counted by $N_t^0$.

The operators $N_s^k$, $s\leq t$, $k=0,1$, are indeed the operators that generate the observer algebra $\mathcal{C}_t$, but they are not actual observations because the latter should be given by orthoprojectors $\Pi_{\upsilon^t}$ on $[0,t)$, adapted to $[0,r)$, having the form
\begin{equation}
\Pi_{\upsilon^t}(\vartheta)=I(\vartheta_{[t})\otimes P_{k_n}(t_n)\otimes\cdots P_{k_1}(t_1)\label{piob}
\end{equation}
for the observations $\upsilon^t=\{k_1,\ldots,k_n\}$, $k_i\in\{0,1\}$,  supported on a particular chain $\vartheta^t=\{t_1<\cdots<t_n\}$, and zero otherwise; which necessarily
means that $|\vartheta^t|=|\upsilon^t|$.
 Further, any such orthoprojector that results in $\Pi_{\upsilon^t}\Psi_t=0$ should be regarded as impossible, and only the remaining ones correspond to possible observations. These possible observations
generate the projective subalgebra $P_t\mathcal{C}_t\subseteq\mathcal{C}_t$ and any true
observable should be regarded as a function of the observations that could actually be made. Further, each such possible observation will generally carry with it a different expectation of a quantum property $X$ of the atom. That is an expectation
conditioned by an observation.

\section{Observation and Filtering}\label{fil}
If observables are regarded as questions asked - regarding what might happen, then an actual observation provides a set of answers to such questions.  So the number
operators $N^k_t$ would be asking questions about how many times we observe an event of a particular kind, but they also play a more fundamental role. They support the
stochastic dynamics of the expectation $\epsilon_t(X)$ of a quantum property $X$ conditioned on the set of all possible observations $\Pi_{\upsilon^t}$ given by (\ref{piob}). That is to say that the two number
operators are the random noises that drive the quantum non-linear filtering of quantum information from the atom to the cat. Which means $\epsilon_t(X)$ should evolve according to a %classical
stochastic differential equation of the form
\begin{equation}
\mathrm{d}\epsilon_t(X)=\kappa_{t,k}(X)\mathrm{d}N^k_t\label{sde}
\end{equation}
where  $\mathrm{d}N^k_t$ are the stochastic counting increments
%\footnote{We could have asked a more involved question having more outcomes than the 0 or 1 considered here, for example distinguishing between the cat being alive and being dead. This would mean having a bigger dilation $S$ and having more degrees of freedom in $\mathfrak{k}$. If we have an apparatus that can ask an atom more precise and detailed questions then that is fine. This in turn would  give us more number operators, or even other kinds of observations. For example the apparatus might make a continuous trajectory corresponding to our observation of such a trajectory, and its path would answer other questions we have about the atom (or whatever else we are questioning). The only thing that limits us is our interpretation of the measurement outcomes in relation to what is being questioned. This is probably why physics is a much stronger science than psychology.}
satisfying $\mathrm{d}N^i_t\mathrm{d}N^j_t= \delta^{ij}\mathrm{d}N^j_t$; in (\ref{sde})
Einstein summation convention has been implemented on $k$. The functions $\kappa_{t,k}(X)$ are the stochastic derivatives of $\epsilon_t(X)$ and shall be derived in due course.

There are some important points to make about equation (\ref{sde}). First of all, $\epsilon_t(X)$ is essentially a classical process because it commutes with all operators in $\mathcal{A}_t$, in fact $\epsilon_t:\mathcal{A}_t\mapsto\mathcal{C}_t$. Secondly, $\epsilon_t(X)$ is certainly required as an important concept in quantum mechanics because it  is the only thing that describes the evolution of a quantum property, $X$, \emph{given} all the possible observations that an observer can make. This means that if you want to make inference about the source of your observations then you can only do it from within your observation. Note that this still allows a `many worlds' interpretation of quantum mechanics as a `holistic' perspective because the entangled state $\Psi_t$ is the real physical object at work here. However, from \emph{within} the whole quantum system the observer forms only a part, and your  experience (forming your reality) is but a single path of observation\footnote{In a more complicated model one might suppose that their experience is composed of many sub-observations. In such a case there may indeed be incompatibility between the different sub-observations (for example you might think of each string of observation coming from each atom within your body as a sub-observer from which you are made, and perhaps these observations are not compatible with each other even if they are compatible with themselves and the underlying source). However, if that were so then there would be a breakdown in one's attempt to make predictions about future events based on prior observations. Although practically inconvenient, it may be a more realistic model of our experience of reality.} and only a projection of the universe.

%A continuous observation of Schr\"odinger's cat would correspond to a continuous product of projectors. Here we shall simply consider measurement instants occurring
%anywhere on a time continuum.
So far we have considered the fundamental observables associated with this atom-cat model and evaluated their expected values up to a time $t$ in the future without any
conditioning from actual observations. Those expectations were of course given by the entangled pure state $\Psi_t$. %That is the expected outcome of a specific observation - counting events.
 We have also established that an {actual} observation  is an orthoprojector on the apparatus space corresponding to an
 sequence of observations\footnote{Or measurement readings if one prefers the context of laboratory experiments.} at %each time in an interval $[0,t)$ now identified with the past so that $t$ is  the present time for the observer.
arbitrary instants, $\vartheta^t$.
%Such an observation is a trajectory in the observer algebra.
 Once the observer starts making these actual observations the predictions of any future quantum property in $\mathcal{A}_t$ will of course be conditioned by the past observations.
  Recall that the expectation of the observable counting
 process $N_t^1$ was an increasing function of $t$ going from $0$ up to $|\beta|^2$, the probability that the atom was initially in its excited state.
  But we shall soon see that the \emph{conditional} expectation of $N_t^1$ is precisely
 either 0 or 1 depending on which of only two possible types of  observation are made (excluding `no observation').

 With actual observation in mind we now re-consider our prepared quantum state $\Psi_t^{} \Psi_t^\ast$ entangling the atom and the apparatus on $[0,t)$.
 % and we shall  project this operator %on $\mathscr{H}$
 % onto the subalgebra $\mathcal{A}_t=\mathcal{B}_{[t}\otimes\mathcal{C}^t\subset\mathcal{B}(\mathscr{H})$ % generated by a set of compatible observables of the apparatus.
 % where $\mathcal{B}_{[t}=\mathcal{B}(\mathscr{H}_{[t})$. The entangling operator  $U_t$ %and this may be adapted to an interval of time of arbitrary size and
 % is generally understood as a unitary quantum stochastic process; %$U^t$ in $\mathcal{A}$.
 %  it is a stochastic propagator. In this case it is a quantum counting process and we shall now redefine it as a classical counting process - this is how we
  We would now like to project out a counterpart of this pure quantum state onto $\mathcal{A}_t$
  %In the end we shall see that when all histories are considered the expectation of $X$ is unaltered.
to obtain a new `state'  $\psi(t) \psi(t)^\ast$ which is
  an $\mathcal{A}_t$-valued
  stochastic operator corresponding to a  stochastic evolution of the atomic quantum state $|\psi\rangle\langle\psi|$ driven by random observations.
  %\begin{equation}
  %\Psi(t)^\ast \big(X\otimes \Pi\big)\Psi(t)=\psi(t)^\ast \big(X\otimes \Pi\big)\psi(t)\qquad\forall\;\Pi\in\mathcal{C}.
  %\end{equation}
  As this suggests, the central object of study here % in the quantum non-linear filtering theory:
  is the  stochastic `wave-function' $\psi(t)$, which  lives in the Hilbert $\mathcal{C}_t$-module $\mathfrak{H}_t$,  as
  it  defines the conditional expectation  as a linear map $\epsilon_t:\mathcal{A}_t\mapsto \mathcal{C}_t$  on the basic elements $A=X\otimes B_{[t}
  \otimes C^t$ as
  \begin{equation}
  \epsilon_t(A)={c}_t\psi(t)^\ast X\psi(t),
  \end{equation}
  where $c_t=\Phi_{[t}^\ast B_{[t}^{}\Phi_{[t}^{}C_t=C_t$ if $B=I$. Note   that $\mathcal{C}_t=I_{[t}\otimes\mathcal{C}^t=\mathcal{A}_t^\prime$.
  This is the \emph{quantum non-linear filtering} of any quantum property $X\in\mathcal{B}(\mathfrak{h})$.
   This means that this conditional expectation corresponds to the filtering of information, about the atom, onto the observer's algebra.
   To  determine exactly what $\psi(t)$ is will depend on what $\mathcal{C}_t$ is, i.e. what kind of observations are being made, and we proceed as follows.
 % The result of this projection this projected propagator $V(t)$. In doing this we lose the explicit appearance of the apparatus state
 % $|\Phi\rangle\langle \Phi|$ and we also lose the unitarity of the evolution. The former is no problem, for the information held by the initial apparatus state becomes absorbed into
 % $V(t)$ and the latter can be resolved by simply re-normalizing; although that does give rise to a dependence on the quantum state $\psi$. The resulting operator
 % $V(t;\psi)$ defines a filtration - that is the filtering of information from the atom onto the observer's algebra - the space of actual observations. The projection of
 % the unitary quantum stochastic evolution  onto $\mathcal{C}$ defines the conditional expectation $\epsilon:\mathcal{A}\rightarrow\mathcal{C}$ given on the
 % quantum properties $X$ as %cond. expec. equn.
 %  where $\psi(t)=V(t;\psi)\psi$ lives in the module $\mathcal{H}=\mathfrak{h}\otimes\mathcal{B}(\mathscr{F})$ and the exact form of $V$ will depend on exactly what
 % kind of observations are made as we shall now see.
%
%First we shall first consider counting-type observations which are quite straightforward.
%

First we shall recall the observation projectors $\Pi_{\upsilon^t}$ where $\upsilon^t$ is a
 sequence  of specific outcomes, say $1,0,0,1,0,1,1,0,0,0...$, supported on a  chain $\vartheta^t$ indicating the instants of these outcomes.
 For this particular atom-cat model there are only three possible basic observations and they form a set of orthogonal orthoprojectors. The first is `no observation' corresponding to the adapted vacuum projector $\Pi_t^\emptyset(\vartheta)=I(\vartheta_{[t})\otimes O(\vartheta^t)$, where $O(\vartheta)=1$ if $\vartheta=\emptyset$ and zero otherwise. The other two correspond to non-trivial observations and they are
   represented by the orthoprojectors
\begin{equation}
\Pi_t^0(\vartheta)=I(\vartheta_{[t})\otimes P_0^\otimes(\vartheta^t)\quad\textrm{and}\quad
\Pi_t^1(\vartheta)=I(\vartheta_{[t})\otimes P_0^\otimes(\vartheta^t\setminus t_1)\otimes  P_1(t_1),
\end{equation}
as any non-zero projection $\Pi_{\upsilon^t}(\vartheta)\Psi_t(\vartheta)\neq0$ necessarily has $\Pi_{\upsilon^t}(\vartheta)$ equal to $\Pi_t^0(\vartheta)$ or $\Pi_t^1(\vartheta)$
and any other observation is simply impossible: $\Pi_{\upsilon^t}(\vartheta)\Psi_t(\vartheta)=0$. Note that the vacuum values of these two projectors are
$\Pi_t^k(\emptyset)=0$, and that may be taken as definition.

Next we project the pure quantum state $\Psi_t^{}\Psi_t^\ast$ onto $\mathcal{A}_t$ defining % we define the filtering density operator  as follows.
  \begin{equation}
  \pi(t):= \sum_{\upsilon^t} \Pi_{\upsilon^t} U_t\Big(|\psi\rangle\langle\psi|\otimes\Phi\Phi^\ast\Big)U_t^\ast\Pi_{\upsilon^t}\equiv V(t) \Big(
  |\psi\rangle\langle\psi|\otimes I\Big)V(t)^\ast,\label{pi}
  \end{equation}
  %where $\mu=\nu^\otimes e^{-\nu t}$ is the Fock space representation of a Poisson measure.
  %where $\mu=\nu^\otimes e^{-\nu t}$ is the Fock space representation of a Poisson measure.
    where $V(t)=\sigma^\odot_t\sqrt{\nu}^\otimes e^{-\frac{1}{2}\nu r}$ and
  \begin{equation}
  \sigma_t^{\odot}(\vartheta):=I(\vartheta_{[t})\otimes\left(
            \begin{array}{cc}
              JJ^\dag & 0 \\
              0 & J \\
            \end{array}
          \right)^{\odot}(\vartheta^t),
  \end{equation}
   noting that $\texttt{Tr}[\pi(t)A]=\Psi_t^\ast A\Psi_t^{}$ for any $A\in\mathcal{A}_t$.
  This may be regarded as the first step to turn the quantum stochastic pure-state $\Psi_t^{}\Psi_t^\ast$ into a causal structure that allows predictions to be made from observations. The summation in (\ref{pi}) is in fact also a direct integral over all chains $\vartheta$, but this has been dropped from the
  notation for brevity.
  Notice that the apparatus state $\Phi$ has now been absorbed into the new propagator $V(t)$ and
  of course the evolution must be diagonal as we are working in the diagonal representation of the apparatus.
  %Also note that  $[\pi(t),\Pi_{\upsilon^t}]=0$ as one would expect for all operators in $\mathcal{C}_t$.

We are not yet finished in our construction of the filtering wave-function as the propagator $V(t)$ is not unitary. Instead, we would like to have a `quasi-unitary' operator, $V_\psi(t)$ say, so that
  \begin{equation}
      \pi(t)=\sum_{\upsilon^t} \Pi_{\upsilon^t} U_t\big(|\psi\rangle\langle\psi|\otimes\check\varrho\big)U_t^\ast\Pi_{\upsilon^t}= V_\psi(t) \big(
  |\psi\rangle\langle\psi|\otimes \hat\varrho_t\big)V_\psi(t)^\ast,\label{pi2}
  \end{equation}
  where $\check\varrho=\Phi\Phi^\ast=(\nu P_0)^\otimes e^{-\nu r}$ is the \emph{input} distribution (the `prepared' state of the apparatus) and $\hat\varrho_t$ is the \emph{output} distribution containing the probabilities of the different observations that can be made.
The operator $V_\psi(t)$ may be obtained by a re-normalising procedure which is where the non-linearity of the filtering comes from. So, finally, %we impose a normalization
   %of $\pi(t)$ %when we trace out $\mathfrak{h}$ leaving the identity in $\mathcal{B}(\mathscr{F})$.
   %This will make the re-normalized operator $V(t;\psi)$ unitary operator on $\mathcal{H}$.
   to achieve this  we must invert
   $\texttt{Tr}[\pi_\upsilon(t)]$ for all non-zero $\pi_\upsilon(t):=\Pi_\upsilon\pi(t)$. This immediately
  gives us the  quasi-unitary operator $V_\psi(t)$ on the subspace of possible observations having the $\vartheta$-evaluations that are generally
  no longer of decomposable semi-tensor product form, but instead
  \begin{equation}
  V_\psi(t,\vartheta)=\sigma_\psi(t_n,\vartheta)\cdots\sigma_\psi(t_2,\vartheta)\sigma_\psi(t_1,\vartheta)
  \end{equation}
  where $\sigma_\psi(t,\vartheta)=I(\vartheta_t)\otimes\sigma_\psi(t,\vartheta^t)$. But remember, this model is a very simple special case, so we shall now see that the decomposable tensor product structure is preserved for us here. First, we shall
  explicitly consider $\sigma_\psi(t_2,t_1)$, for example, it would generally have the form
 \begin{equation}\left[
                                   \begin{array}{cc}
                                      \left(
            \begin{array}{cc}
              JJ^\dag/\|J^\dag\psi_0\| & 0 \\
              0 & J/\|J\psi_0\| \\
            \end{array}
          \right) & 0 \\
                                     0 &  \left(
            \begin{array}{cc}
              JJ^\dag/\|J^\dag\psi_1\| & 0 \\
              0 & J/\|J\psi_1\| \\
            \end{array}
          \right) \\
                                   \end{array}
                                 \right],\label{bloc}
  \end{equation}
  where the wave functions $\psi_0:=\big(JJ^\dag/\|J^\dag|\psi\rangle\|\big)|\psi\rangle$ and $\psi_1:= \big(J/\|J|\psi\rangle\|\big)|\psi\rangle$ simply refer to the two possible outcomes from the first interaction given by $\sigma_\psi(t_1)$.

  Next, note that $\|J\psi_0\|$ and $\|J\psi_1\|$ are both zero, no matter what $\psi_0$ and $\psi_1$ are, so those terms simply don't contribute to the evolution and are removed - those observations are not possible,
   and since $\|J^\dag\psi_0\|$ and $\|J^\dag\psi_1\|$ are both 1 we  get
  \begin{equation}
  \sigma_\psi(t_2,t_1):=                              \left(
            \begin{array}{cc}
              JJ^\dag & 0 \\
              0 & 0  \\
            \end{array}
          \right)\otimes Q(t_1),
  \end{equation}
  where $Q=I$ if neither $\alpha$ nor $\beta$ are zero, $Q=P_0$ if $\beta=0$ and $Q=P_1$ if $\alpha=0$, recalling that $|\psi\rangle=\alpha|g\rangle +\beta|e\rangle$.
  So in this very simple model the decomposable tensor product structure is preserved and a non-trivial re-normalisation will only occur at the first instance of an
  interaction, i.e for $\sigma_\psi(t_1)$. This can be seen very easily from (\ref{Psi}). Further, we find that the output distribution $\hat\varrho_t=\hat\varrho_{[t}\otimes\hat\varrho^t$ has
   \begin{equation}
      \hat\varrho^t=e^{-\nu t}\big(1\oplus\nu\rho\oplus\nu^2\big(P_0\otimes\rho\big)\oplus\ldots\big),\qquad\rho= \left(
            \begin{array}{cc}
              |\alpha|^2 & 0 \\
              0 & |\beta|^2  \\
            \end{array}
          \right)
      \end{equation}
and $\hat\varrho_{[t}=\check\varrho_{[t}= (\nu P_0)^\otimes e^{\nu(t-r)}$.

Now we have the $\mathfrak{H}_t$-valued filtering wave-function   given explicitly as $\psi(t)=V_\psi(t)\psi$ with $\psi(t,\vartheta)= |g\rangle\otimes \big(\Pi^0_t(\vartheta)+\Pi^1_t(\vartheta)\big)$ except $\psi(t,\emptyset)=|\psi\rangle$, satisfying the condition
\begin{equation}
\epsilon_t(I)=\psi(t)^\ast\psi(t):=P_t=\sum_{\upsilon^t}\Pi_{\upsilon^t}
\end{equation}
where the sum is only taken over possible observations (ones with non-zero probability) which in this case simply means $P_t=\Pi_t^\emptyset+\Pi_t^0+\Pi_t^1$. Indeed $P_t\Psi_t=\Psi_t$ and $\epsilon_t(A)P_t=\epsilon_t(A)$ for all
$A\in\mathcal{A}_t$.
 Also note that $\epsilon_t(\Pi_{\upsilon^t})=\Pi_{\upsilon^t}$ for possible observations, otherwise it is zero, which really just reflects the fact
 that $\epsilon_t(C)=P_tC$ for all $C\in\mathcal{C}_t$. Ultimately we see that quantum properties can be conditioned on the three mutually exclusive events $\Pi_t^\emptyset$, $\Pi_t^0$ and $\Pi_t^1$ such that
 \begin{equation}
 \mathbb{E}[A]=\mathbb{E}[A|\Pi_t^\emptyset]\mathbb{E}[\Pi_t^\emptyset]+\mathbb{E}[A|\Pi_t^0]\mathbb{E}[\Pi_t^0]+ \mathbb{E}[A|\Pi_t^1]\mathbb{E}[\Pi_t^1].
 \end{equation}

  We may easily define  $\mathfrak{h}$-valued wave-functions $|\psi_{\upsilon^t}\rangle$ by identifying $\Pi_{\upsilon^t}$ with $\Upsilon^{}_t\Upsilon_t^\ast$ so that
   $|\psi_{\upsilon^t}\rangle:=\Upsilon_t^\ast\Psi_t/\|\Upsilon_t^\ast\Psi_t\|$
   %for the counting-type observations %\sigma^\odot(\upsilon^t)\psi/\|\sigma^\odot(\upsilon^t)\psi\|$
   are the evolutions of $|\psi\rangle$ resulting from specific observations $\upsilon^t$.
   %Note that this is a familiar axiom of quantum measurement.
  %resulting from the observation $\upsilon^t$.
  This allows us to write $\Pi_{\upsilon^t}\psi(t)=|\psi_{\upsilon^t}\rangle\otimes\Pi_{\upsilon^t}$ so that the conditional expectation may be given on
  any $X\in\mathcal{B}(\mathfrak{h})$ in the forms
  \begin{equation}
  \epsilon_t(X)= \sum_{\upsilon^t}\left( \frac{\Psi_t^\ast\big( X\otimes\Pi_{\upsilon^t}\big) \Psi^{}_t}{ \Psi_t^\ast\Pi_{\upsilon^t} \Psi_t^{}}\right)
   \Pi_{\upsilon^t}
  =\sum_{\upsilon^t} \langle\psi_{\upsilon^t}| X |\psi_{\upsilon^t}\rangle  \Pi_{\upsilon^t}\label{geneq}
  \end{equation}
  which illustrates that each conditional expectation is carried on the appropriate projector representing an actual observation, and recall that such a projector represents a trajectory of the
  apparatus: a specific observation of the cat. Thus we have different possible histories that may arise in any given experiment and each one carries, in principle, a different expectation of the property $X$. Do bear in mind that $\langle\psi_{\upsilon}| X |\psi_{\upsilon}\rangle =\mathbb{E}[X|\Pi_{\upsilon}]$ and the quantum causality condition means that $\mathbb{E}[X]=\sum_{\upsilon}\mathbb{E}[X|\Pi_\upsilon]\mathbb{E}[\Pi_\upsilon]$ which requires that $[X,\Pi_\upsilon]=0$. %Notice that we can condition quantum properties on observations, but we cannot condition future observations on quantum properties,  only on past observations, e.g. $\mathbb{E}[\Pi_{\upsilon^t}|\Pi_{\upsilon^s}]$.

  %and for the purposes of direct computation one may realize that the wave-function $\psi_\upsilon = \langle \Upsilon^t|\Psi(t)/\|\langle \Upsilon^t|\Psi(t)\|$ so that the conditional
  %expectation may be given entirely in terms of the quantum pure state $\Psi$. This follows simply from the correspondence of $\Pi_{\upsilon^t}$ with
  %$ |\Upsilon^t\rangle \langle \Upsilon^t|$ for the counting-type observations.

  Let's now recall the actual form of the possible observations here. There were only three.
  %From the above it follows that there are two possible observations on any chain $\vartheta$.
  On any chain $\vartheta$ we either have $\Pi_t^\emptyset(\vartheta)$, $\Pi^0_t(\vartheta)$ or $\Pi^1_t(\vartheta)$ respectively corresponding to $\upsilon^t=\emptyset$,
  $\upsilon^t=0000000\ldots$ and $\upsilon^t=1000000\ldots$ (although we can indeed consider different chains as different observations, i.e. different instants when the individual events occur).
  %And when we write $\Pi_{\upsilon^t}$ we mean that for the $\vartheta^t$ supporting $\upsilon^t$ it is either $\Pi_t^0(\vartheta^t)$ or $\Pi_t^1(\vartheta^t)$.
  %The first of these is the survival process $\Pi_0$ from the previous section,
  First let's see if an observation of the form $\Pi_t^\emptyset$ will occur. Since
  \begin{equation}
\mathbb{E}_t[\Pi^\emptyset_t]=e^{-\nu t}
\end{equation}
  we see that the possibility of this observation rapidly decays to zero.
 Next we shall see whether or not an observation of the form $\Pi^0_t$ will occur and the answer that we get back will look like
\begin{equation}
\mathbb{E}_t[\Pi^0_t]=(1- e^{-\nu t})|\alpha|^2\label{decc}
\end{equation}
which rapidly approaches 1 if the atom begins in the ground state, but if the atom begins in the excited state this observation will not happen because the atom will decay and the cat will die.  More specifically, (\ref{decc}) says that as $t$ gets large we would only expect to have an observation of the form $\Pi_t^0$ with probability $|\alpha|^2$. So the more dominant the ground-state contribution to the initial atomic state $|\psi\rangle$ the more likely it is that a $\Pi_t^0$ observation will take place. After all, if the cat was not observed to die then the atom is likely to have initially have been in its ground state $|g\rangle$.

  The final type of observation, $\Pi^1_t$,
   corresponds to the indirect observation (i.e. inference) of the decay of the atom by observing the cat's death. The expectation is
  \begin{equation}
\mathbb{E}_t[\Pi^1_t]=(1- e^{-\nu t})|\beta|^2=\mathbb{E}_t[N^1_t],\label{dd}
\end{equation}
and note that $\mathbb{E}_t[\Pi^\emptyset_t+\Pi^0_t+\Pi^1_t]=1$. So this observation will not happen if the atom is initially in ground-state, but otherwise it will eventually happen with probability $|\beta|^2$. Both (\ref{decc}) and (\ref{dd}) demonstrate that the atomic state will determine what observations will take place.
%which we shall see follows from the fact that
An important point to note is that  both of these observations  will transform the atomic wave-function  to $|g\rangle$, either by $\ldots F_0 F_0|\psi\rangle$ or $\ldots F_0 F_1|\psi\rangle$. Whereas the `non-observation' $\Pi_t^\emptyset$ will have no effect on $|\psi\rangle$.

Since any operator $C\in\mathcal{C}_t$ satisfies $\epsilon_t(C)=P_tC$
we have $\epsilon_t(N^k_t)=P_tN^k_t$ which simply projects the number operators onto that part of the observer algebra generated by possible
observations. For  $N^1_t$ we find that its projection gives us
\begin{equation}
\epsilon_t(N^1_t)=\Pi_t^1,
\end{equation}
which we may now interpret as a restricted number operator that only counts either 0 or 1, depending on the observation $\Pi_t^k$.
The part of $N^1_t$ that counts above 1 is observationally impossible and has an equivalence with zero as it has no part overlapping
with $\Psi_t$. Indeed $\Pi^1_t$ is truly an observable counting  the death of Schr\"odinger's cat. %So, as it turns out, the only possible observations one can
%make here have an intimate relationship with
%More generally we can see that $\epsilon_t(C)=CP_t$ for all observables $C$ in $\mathcal{C}_t$.
In the case of $N^0_t$ we find that
\begin{equation}
\epsilon_t(N^0_t)=n_t\Pi_t^0+(n_t-1)\Pi_t^1,
\end{equation}
where $n_t(\vartheta):=|\vartheta^t|$, which results in a count of $n_t$ if there is an observation of type $\Pi_t^0$ and a count of $n_t-1$ if there is on observation of type $\Pi_t^1$, as we would expect.

Now the conditional expectation of a quantum property $X\in\mathcal{B}(\mathfrak{h})$ may be defined by its
%assuming that its evaluation on the empty chain is zero? Yes - consider creation op. acting on product state...
evaluations  on the chains $\vartheta$ as
  \begin{equation}
  \epsilon_t(X,\vartheta)= \langle g| X| g\rangle \big(\Pi_t^0(\vartheta)+\Pi_t^1(\vartheta)\big)\label{a}
  \end{equation}
  for $\vartheta^t\neq \emptyset$.   Otherwise, when $\vartheta^t=\emptyset$,  there is no observation so $\Pi_{\upsilon^t}=\Pi_t^\emptyset$, the vacuum projector, and
  \begin{equation}
  \epsilon_t(X,\vartheta) =\langle\psi| X|\psi\rangle\Pi_t^\emptyset(\vartheta).\label{b}
  \end{equation}
 Then, since $\Pi_t^0+\Pi_t^1$ is zero on $\emptyset$ and $\Pi_t^\emptyset$ is zero unless $\vartheta^t=\emptyset$, one may simply write
  \begin{equation}
  \epsilon_t(X)=\langle \psi|X|\psi\rangle\Pi_t^\emptyset+\langle g|X|g\rangle \big(\Pi_t^0+\Pi_t^1\big),
  \end{equation}
  and note that the filtering wave-function is $\psi(t)=|\psi\rangle \otimes \Pi_t^\emptyset +|g\rangle\otimes (\Pi_t^0+\Pi_t^1)$.
  Indeed it is easy to see from the definition of $\epsilon_t$ in terms of $\Psi_t$ (\ref{geneq}) that
  \begin{equation}
  \mathbb{E}_t\big[\epsilon_t(A)\big]=\mathbb{E}_t\big[A\big]\label{ce}
  \end{equation}
  and so it is of no surprise that  for $A=X\otimes I$ we have
  \begin{equation}
 \mathbb{E}_t\big[\epsilon_t(X)\big]= \langle \psi| X|\psi\rangle  e^{-\nu t} + \langle g| X|g\rangle  \big(1-e^{-\nu t}\big),
  \end{equation}
  which is of course $\texttt{Tr}[X\rho(t)]$ with $\rho(t)$ given by (\ref{1}). % So notice that the semi-classical representation of the quantum pure-state in the Hilbert
  %$\mathcal{C}_t$-module does not have any affect on the expectation of a quantum property.
  % \begin{equation}
  %\epsilon(X)= \psi^\dag X\psi\otimes \delta_\varnothing + \mu\Pi_0\psi^\dag_g X\psi_g%\int^\oplus\mathrm{d}\mu(\vartheta) P_0^\otimes(\vartheta) \psi^\dag_g X\psi_g
  %+ \mu\Pi_1 \psi^\dag_g X\psi_g
  %\end{equation}
  %Notice that we have assumed that both $\|J\psi\|$ and $\|J^\dag\psi\|$ are non-zero.
  %If $\|J\psi\|=0$ then everything here is completely trivial, but if $\|J^\dag\psi\|=0$ then the atom is initially in its excited state and the generator of the
  %filtering synamics is simply
  %\[
  %\sigma(\psi)=J
  %\]
 In order to find the incremental coefficients $\kappa_{t,k}(X)$ of (\ref{sde}) we can use the filtering wave-function derivatives since $\epsilon_t(X):=\psi(t)^\ast X\psi(t)$, so that
 \begin{equation}
 \kappa_{t,k}(X)=\langle k|\big(\psi(t)^\ast\sigma_\psi(t)^\ast X\sigma_\psi(t)\psi(t)-\psi(t)^\ast X\psi(t)\big)|k\rangle,
 \end{equation}
then we find that if $t=t_1$
 \begin{equation}
  \kappa_{t,k}(X,\vartheta)= \big(\langle g|X| g\rangle-\langle\psi|X|\psi\rangle\big)I(\vartheta).
  \end{equation}
Otherwise
    \begin{equation}
  \kappa_{t,0}(X,\vartheta)=0,\quad\textrm{and}\quad  \kappa_{t,1}(X,\vartheta)=-\epsilon_t(X,\vartheta),
  \end{equation}
which tells us that counting events of type-0 does not change the conditional expectation and counting events of type-1 returns zero.

Finally we shall consider the conditional expectations of the `cat death' counting operator $\Pi^1_t$ carried on the orthoprojetors $\Pi_{\upsilon^t}$. These are simply given by
$\Pi^1_t\Pi_{\upsilon^t}$ and are either $0$,  which arise
whenever $\Pi_{\upsilon^t}(\vartheta)=\Pi^0_t(\vartheta)$, or they are $1$, corresponding to cases when $\Pi^1_t\Pi_{\upsilon^t}=\Pi_{\upsilon^t}$
 which arise
whenever $\Pi_{\upsilon^t}(\vartheta)=\Pi^1_t(\vartheta)$. Thus for any possible observations that may occur here the expectations of $\Pi^1_t$, and
indeed $N^1_t$, \emph{conditioned}
on such observations are exactly 0 or 1, the latter arising if and only if the death of the cat is observed.

Thus we have established a fundamental classical stochastic process $\epsilon_t(A)$ for this model. It is well defined for any $A\in\mathcal{A}_t$ and its trajectories are the
conditional expectations of $A$ combined with the corresponding projectors. A historical string of observations  is  one such trajectory and may be obtained for any system property $X$
 as
\begin{equation}
\Pi_{\upsilon^t}\epsilon_t(X)=\langle\psi_{\upsilon^t}|X|\psi_{\upsilon^t}\rangle \Pi_{\upsilon_t}
\end{equation}
and this is zero unless $\upsilon^t=\emptyset$, $000\ldots$, or $100\ldots$, respectively resulting in $|\psi_{\upsilon^t}\rangle=|\psi\rangle$, $|g\rangle$, or $|g\rangle$. This
means, for example, that if we ask the question: \emph{``is the atom excited?''}, corresponding to $X=|e\rangle\langle e|$, then any non-empty observation will return the certain answer \emph{``no''}. This answer seems reasonable if we observed the cat's death, i.e. an observation of the form $\Pi^1_t$, but it seems rather unsatisfactory if our observation was of the form $\Pi^0_t$. However, on closer inspection we realise that if our observation is of the form $\Pi_t^0$ then the cat has not died, and since we assume it was alive to begin with this means that the atom has not decayed. Further,  this model does not allow events of type 1 to follow events of type 0, so that means if we have the observation $\Pi_t^0$ we know that the cat won't die at a later time and so the atom must have initially been in the ground state. This is not Zeno's paradox at work, it is simply due to  the simplistic choice of the interaction operator (\ref{s}). In the sequel we shall see that a description of the cat using three degrees of freedom will resolve this issue and give a very satisfactory model of Schr\"odinger's Cat. That is not discussed here because this article has become quite long enough.

%However, in the latter case we must concede that certain properties of the construction here. First of all the atom has been assumed to decay, by construction, and it has also been assumed to be coupled to the cat in such a way that we have \emph{decay} $\Leftrightarrow$ \emph{cat dies}. The problem here is that  observation $\Pi^0_1$ corresponds to the action of $|g\rangle \langle g|$ on the atom.

\section{Conclusion}
Let's recall the main points that have been discussed here. One experiences sensation which shall basically be referred to as observation, in some general sense. This observation is then believed to have a source - something that gives rise to it. Given the apparent quantum nature of things the source and the observation are then considered to be two quantum systems coupled together. Both the source and the observation are represented in their own Hilbert space, each in some relevant basis. Operators associated with the source are quantum properties not observables  and they are built from arbitrary projectors that need not commute. Operators associated with the observation are observables and they are built from commuting projectors which also commute with quantum properties. If this structure is dismissed then the observations have no predictive structure. Consequently, the possible observations form a commutative subalgebra within the whole algebra of the source-observation quantum system. Information about the source can be projected onto the algebra of the observations and this is the filtering of quantum information by the observation. And that is also the description of the possible observations that can arise.

Belavkin had called this quantum filtering  ``Eventum Mechanics'' by which `event enhanced quantum mechanics' was meant. It can be formulated in the Schr\"odinger picture as a future-past boundary value problem in which `incoming' quantum information from the future gives rise to `outgoing' classical\footnote{Classical does not mean `position and momentum commute' or anything like that, it means that the information forms a commutative algebra.} information in the past. It may sound counterintuitive that the information travels `backwards in time', i.e. from future to past, but it is actually quite reasonable. Consider your next meal, where is it? in the past or future? It's in the future, and it is approaching, getting closer and closer. When you eat it will have arrived, and when you finish that meal it will have gone into the past.

One can argue that observation is being generated from `nothing', it just happens, but in a quantum framework that can still be described by a quantum source. In fact it is more interesting for the source and the observation to have the same mathematical structure (namely Guichardet-Fock space) because it gives a duality to the system. That is the whole is composed of two coupled subsystems who are generating one another's observations. Even in the atom-cat system considered above it is evident that the observation of the cat is as responsible for the fate of the atom as the atom is for the fate of the cat. So it would seem reasonable that if this toy `universe' were turned inside out then the atom would serve as the observation and the cat the source. But beware! in this dual universe the atom would have to be described by a commutative subalgebra whilst the cat would become fully quantum.

\section{Appendix Part I}\label{ap1}
The Guichardet-Fock space $\mathscr{F}$ over $[0,r)$ is a Hilbert space of functions $\varphi$ which may be defined in terms of their evaluations $\varphi(\vartheta)$
on finite chains $\vartheta\subset[0,r)$. These evaluations are themselves elements of the appropriate product Hilbert space $\mathfrak{k}^{\otimes|\vartheta|}$
where $|\vartheta|$ is the cardinality of $\vartheta$. In this way one may then consider $\varphi$ as the direct integral
\[
\varphi=\int^{\oplus}\mathrm{d}\vartheta \varphi(\vartheta)
\]
which may be written in Dirac notation if one prefers, but that can run into issues with $\delta$-functions when trying to define $P_t$. The integral $\int\mathrm{d}\vartheta$ is shorthand
for $\sum_{n=0}^\infty \int\mathrm{d}\vartheta_n$ where $\vartheta_n$ is any chain of cardinality $n$ and $\int\mathrm{d}\vartheta_n
=\int_0^t\ldots\int_0^{t_2}\mathrm{d}t_1\ldots\mathrm{d}t_n$ is the chain integration over a simplex with $\vartheta_n=\{t_1<\cdots<t_n\}$.
 The norm of such functions is given by integrating  over all
values of $\vartheta$ so that
\[
\|\varphi\|^2=\int\mathrm{d}\vartheta \|\varphi(\vartheta)\|^2
\]
where $\|\varphi(\vartheta)\|$ is just the norm of $\varphi(\vartheta)$ in $\mathfrak{k}^{\otimes|\vartheta|}$.

One usually considers $\mathscr{F}$
to be the closed linear span of basic product functions $\varphi=\xi^\otimes$ having the evaluations
\[
\xi^\otimes(\vartheta)=\xi(t_n)\otimes\cdots\otimes\xi(t_1)
\]
for any $\vartheta=\{t_1<\cdots< t_n\}$. Indeed each $\xi(z)$, $z\in\vartheta$, lives in a copy of the Hilbert space $\mathfrak{k}$, but $\xi$ is understood to
live in a Hilbert space $\mathcal{K}$ which is taken to be $\mathfrak{k}\otimes L^2[0,r)$. In terms of $\mathcal{K}$ one may understand that $\mathscr{F}=
\Gamma(\mathcal{K})$, which is the `second quantisation' of $\mathcal{K}$ given by the second quantisation functor $\Gamma$. This functor admits the isomorphism
$\Gamma(\mathcal{K}_1\oplus\mathcal{K}_2)\cong\Gamma(\mathcal{K}_1)\otimes\Gamma(\mathcal{K}_2)$ and so $\mathscr{F}\cong\mathscr{F}_{[t}\otimes \mathscr{F}^t$ for
any $t\in[0,r)$ corresponding to the disjunction $[0,r)=[t,r)\sqcup[0,t)$.

The norm of the product state $\xi^\otimes$ is $\|\xi^\otimes\|=\exp\{\frac{1}{2}\|\xi\|^2\}$ and so one may define the so called \emph{coherent states}
as
\[
\Phi=\xi^\otimes e^{-\tfrac{1}{2}\|\xi\|^2}=|0\rangle^\otimes\sqrt{\nu}^\otimes e^{-\tfrac{1}{2}\nu r}
\]
where the latter equality is a consequence of the special case when  $\xi$ is itself separable in $\mathcal{K}$ so that it has the form $\xi=|0\rangle\otimes\sqrt{\nu}$ say, for any normalized vector $|0\rangle \in\mathfrak{k}$
and arbitrary constant function $\sqrt{\nu}\in L^2[0,r)$.

For these purposes it is sufficient to consider only a very special subalgebra of block diagonal operators on $\mathscr{F}$. Such operators $B$ have a multiplication-type
 action
 \[
 [B\varphi](\vartheta)=B(\vartheta)\varphi(\vartheta),
 \]
but the blocks $B(\vartheta)$ may be quite general. A very special case of such operators is when $B$ is a product operator $S^\otimes$ so that
\[
B(\vartheta)=S(t_n)\otimes\cdots\otimes S(t_1).
\]
With respect to the decomposition $\mathscr{F}_{[t}\otimes \mathscr{F}^t$ of $\mathscr{F}$ one may consider the operators of the form $B=B_{[t}\otimes
B^t$ which may, for example, have the form $I_{[t}\otimes
B^t$. In that case the operator is said to be `adapted'. In the case of
an adapted product operator we have
\[
B_t(\vartheta)=I(\vartheta_{[t})\otimes S^\otimes(\vartheta^t)
\]
and if this is to hold for all $t$, considered as an evolution parameter, then $B_t$ is a quantum stochastic Poisson process. In fact $\mathscr{F}$ is the
representing Hilbert space for quantum stochastic processes.

If we now compose $\mathscr{F}$ with another Hilbert space $\mathfrak{h}$ to form $\mathscr{H}=\mathfrak{h}\otimes \mathscr{F}$
then the operators $B_t$ may be considered to have an action on $\mathfrak{h}$ too. In the case of such adapted product operators we may write
\begin{equation}
B_t(\vartheta)=I(\vartheta_{[t})\otimes S^\odot(\vartheta^t)\equiv S^\odot_t(\vartheta)\label{odot}
\end{equation}
where $\odot$ is a tensor product in $\mathscr{F}$ but the usual operator/matrix product in $\mathfrak{h}$. For example, if $S(z)=H\otimes K(z)$ on
$\mathfrak{h}\otimes \mathfrak{k}$, and for $t_1<t_2$, then $S^\odot(t_1,t_2)$ is
\[
S(t_2)\odot S(t_1)=\Big( H\otimes K(t_2)\otimes I(t_1)\Big)\Big(    H\otimes I(t_2)\otimes K(t_1)\Big)
=H^2\otimes K(t_2)\otimes K(t_1).
\]

\section{Appendix Part II}\label{ap2}
Rather than considering the evolution propagator $U_t= S^\odot_t$ transforming the initial state
$|\psi\rangle\otimes\Phi$ one may perform a Weyl transform, given by a unitary operator $W$, on all operators and state-vectors of the system so that the dynamics is represented on the Fock vacuum state
\[
\delta_\emptyset=\left(
                   \begin{array}{c}
                     0 \\
                     0 \\
                   \end{array}
                 \right)^\otimes = W^\ast \Phi,
\]
in which case $U_t$ is transformed into $W^\ast U_t W$. The algebra $\mathcal{A}_t$ is of course transformed accordingly as all elements become $W^\ast A W$ but
since $W$ acts trivially as the identity on $\mathfrak{h}$ it commutes with all quantum properties $X\in\mathcal{B}(\mathfrak{h})$ so that $W^\ast X W=X$.

In order to better understand $W$ we must work in the Belavkin representation of quantum stochastic calculus: Quantum stochastic evolutions are represented as
operators, generally not block-diagonal ones, on $\mathscr{H}$ but Belavkin proved that they may be represented as block-diagonal multiplication operators in a bigger
space $\mathbb{H}=\mathfrak{h}\otimes\mathbb{F}$, where $\mathbb{F}=\Gamma(\mathbb{K})$, and
\[
\mathbb{K}=L^1[0,r)\oplus\mathcal{K}\oplus L^\infty[0,r),
\]
which is a pseudo-Hilbert space when equipped with the Lorentz representation of the Minkowski metric, namely
\[
\boldsymbol{\eta}=\left(
                    \begin{array}{ccc}
                      0 & 0 & 1 \\
                      0 & I & 0 \\
                      1 & 0 & 0 \\
                    \end{array}
                  \right).
\]
With respect to this pseudo-metric we define a $\star$-involution $\mathbf{K}^\star=\boldsymbol{\eta}\mathbf{K}^\ast\boldsymbol{\eta}$ on operators $\mathbf{K}
\in\mathcal{B}(\mathbb{K})$ and on column vectors ${\xi}\in\mathbb{K}$ we have the conjugate rows $\xi^\star=\xi^\ast \boldsymbol{\eta}$.

The pseudo-vacuum, or $\star$-vacuum, state is a product of the physical %gb
gauge vectors
\[
\xi_\emptyset=\left(
                \begin{array}{c}
                  0 \\
                  0 \\
                  1 \\
                \end{array}
              \right)
\]
and the block diagonal operators representing quantum stochastic processes are in fact upper triangular too, so that we are not considering operators in
$\mathcal{B}(\mathbb{K})$
but rather an upper triangular $\star$-monoid resembling  the Heisenberg group; that is indeed invariant under the $\star$-involution.

With all this in mind, the Belavkin representation of $W$ is the operator $\mathbf{W}=\mathbf{Z}^\otimes$ given by the $\star$-unitary operator
\[
\mathbf{Z}=\left(
             \begin{array}{ccc}
               1 & -\xi^\ast & -\tfrac{1}{2}\xi^\ast\xi \\
               0 & I & \xi \\
               0 & 0 & 1 \\
             \end{array}
           \right),
\]
where $\xi=|0\rangle\sqrt{\nu}$. To obtain the transformation of $U_t$ we must first trivially dilate $S$ to form $\mathbf{S}$ and transform it as
$\mathbf{Z}^\star \mathbf{SZ}$ to get
\[
\left(
             \begin{array}{ccc}
               1 & \xi^\ast & -\tfrac{1}{2}\xi^\ast\xi \\
               0 & I & -\xi \\
               0 & 0 & 1 \\
             \end{array}
           \right)\left(
             \begin{array}{ccc}
               1 & 0 & 0 \\
               0 & S & 0 \\
               0 & 0 & 1 \\
             \end{array}
           \right)\left(
             \begin{array}{ccc}
               1 & -\xi^\ast & -\tfrac{1}{2}\xi^\ast\xi \\
               0 & I & \xi \\
               0 & 0 & 1 \\
             \end{array}
           \right)=\left(
             \begin{array}{ccc}
               1 & L^\ast & -\tfrac{1}{2}L^\ast L \\
               0 & S & L \\
               0 & 0 & 1 \\
             \end{array}
           \right)
\]
where $L=(S-I)\xi$, and note that $(S-I)^\ast(S-I)=2(I-S)$ in the decaying atom example considered in this article. Next notice that the resulting $\star$-unitary
operator is equivalent to another $\star$-unitary operator
\[
\mathbf{G}=\left(
             \begin{array}{ccc}
               1 & -L^\ast & -\tfrac{1}{2}L^\ast L \\
               0 & I & L \\
               0 & 0 & 1 \\
             \end{array}
           \right)
\]
on the gauge vector $\xi_\emptyset$. Either way, the object of interest is the $\star$-isometry $\mathbf{F}=\mathbf{G}\xi_\emptyset$ which defines the marginal density matrix
as $\rho(t)=\mathbf{V}_t^\star \big(\rho\otimes\mathbf{I}\big)\mathbf{V}^{}_t$ where $\mathbf{V}_t=\mathbf{F}^{\dag \odot}_t $, which has
the continuous derivative
\[
\frac{\partial}{\partial t}\rho(t)=\mathbf{F}^{\dag\star}\Big(\rho(t)\otimes\mathbf{I}(t)\Big)\mathbf{F}^\dag
\]
which is precisely (\ref{2}) but written here in $\star$-quadratic form.

\section*{Acknowledgements}
A big thanks to Yan Chee Yu for some very heavy duty discussions and disputes on what reality is, and to Vlatko Vedral for testing my belief  in quantum filtering to the point where I even began to doubt it.

\end{document}